# Bifurcations of fractional-order diffusionless Lorenz system

Kehui Sun [a,b*]   J. C. Sprott [b]

[a] *School of Physics Science and Technology, Central South University, Changsha 410083 China*
[b] *Department of Physics, University of Wisconsin-Madison, Madison, WI 53706 USA*

**Abstract**

Using the predictor-corrector scheme, the fractional order diffusionless Lorenz system is investigated numerically. The effective chaotic range of the fractional order diffusionless system for variation of the single control parameter is determined. The route to chaos is by period-doubling bifurcation in this fractional order system, and some typical bifurcations are observed, such as the flip bifurcation, the tangent bifurcation, an interior crisis bifurcation, and transient chaos. The results show that the fractional-order diffusionless Lorenz system has complex dynamics with interesting characteristics.

PACS: 0545

## 1. Introduction

Fractional calculus has a 300-year-old history, as old as calculus itself, but its applications to physics and engineering have just begun [1]. Many systems are known to display fractional-order dynamics, such as viscoelastic systems [2], dielectric polarization, electrode-electrolyte polarization, and electromagnetic waves. Many scientists have studied the properties of these fractional-order systems.

More recently, there has been growing interest in investigating the chaotic behavior and dynamics of fractional order dynamic systems [3-20]. It has been shown that several chaotic systems can remain chaotic when their models become fractional [8]. In [4], it was shown that the fractional-order Chua's circuit with order as low as 2.7 can produce a chaotic attractor. In [5], it was shown that nonautonomous Duffing systems with order less than 2 can still behave in a chaotic manner, and in [6], chaos in a modified Duffing system exists for total system orders are 1.8, 1.9, 2.0, and 2.1. In [7], the fractional order Wien bridge oscillator was studied, where it was shown that a limit cycle can be generated for any fractional order, with an appropriate value of the amplifier gain. In [8], chaotic behavior of the fractional order Lorenz system was studied, but unfortunately, the results presented in that paper are not correct. In [9], the chaotic behavior of a fractional-order "jerk" model was investigated, in which a chaotic attractor was obtained with the system order as low as 2.1, and chaos control of that fractional order chaotic system was reported in [10]. In [11], Chaos in the fractional-order Rössler hyperchaotic system was studied, in which chaos was found in the fractional-order system with an order as low as 2.4 and hyperchaos was found with an order as low as 3.8. In [12], hyperchaotic behavior of an integer-order nonlinear system with unstable oscillators is preserved when the order becomes fractional. In [13-15], the chaotic behavior and its control in the fractional order Chen system are investigated. Many other fractional-order nonlinear systems are chaotic, such as the fractional-order Arneodo's system [16], the fractional-order Chen-Lee system [17], the fractional-order modified van der Pol system [19], and a fractional-order rotational mechanical system with a centrifugal governor [20]. Despite these many examples, the bifurcations of fractional-order nonlinear systems have not yet been well studied.

---
[*] Corresponding author.
*E-mail address:* kehui@csu.edu.cn (K. H. Sun), sprott@physics.wisc.edu (J. C. Sprott)

In this paper, we report the bifurcations that occur in a particularly simple, one-parameter version of the Lorenz model, called the diffusionless Lorenz equations (DLE) described in [21] and further investigated in [22]. The Kaplan-Yorke dimension of the diffusionless Lorenz system was calculated and used as a measure its complexity in [23]. However, the fractional-order variant of this system has not been studied, and it is an ideal candidate for examining bifurcations since it has a single bifurcation parameter. The paper is organized as follows. In Section 2, the numerical algorithm for the fractional-order diffusionless Lorenz system is presented. In Section 3, the chaotic behavior and bifurcations of the system are studied. Finally, we summarize the results and indicate future directions.

## 2. Fractional-order derivative and its numerical algorithm

There are several definitions of fractional derivatives [24]. One popular definition involves a time-domain computation in which non-homogenous initial conditions are permitted and those values are readily determined [25]. The Caputo derivative definition [26] is given by

$$\frac{d^\alpha f(t)}{dt^\alpha} = J^{n-\alpha} \frac{d^n f(t)}{dt^n}, \text{ or } \frac{d^\alpha f(t)}{dt^\alpha} = J^{\lceil\alpha\rceil-\alpha} \frac{d^{\lceil\alpha\rceil} f(t)}{dt^{\lceil\alpha\rceil}}, \qquad (1)$$

where $n$ is the first integer which is not less than $\alpha$ and $J^\theta$ is the $\theta$-order Riemann-Liouville integral operator given by

$$J^\theta \varphi(t) = \frac{1}{\Gamma(\theta)} \int_0^t (t-\tau)^{\theta-1} \varphi(\tau) d\tau, \qquad (2)$$

where $\Gamma(\theta)$ is the gamma function with $0 < \theta \le 1$.

The Diffusionless Lorenz equations are given by

$$\begin{cases} \dot{x} = -y - x \\ \dot{y} = -xz \\ \dot{z} = xy + R \end{cases}, \qquad (3)$$

where $R$ is a positive parameter. When $R \in (0,5)$, chaotic solutions occur. Equation (3) has equilibrium points at $(x^*, y^*, z^*) = (\pm\sqrt{R}, \mp\sqrt{R}, 0)$ with eigenvalues that satisfy the characteristic equation $\lambda^3 + \lambda^2 + R\lambda + 2R = 0$. Figure 1(a) shows the chaotic attractor for the diffusionless Lorenz system for the value of $R = 3.4693$ at which the Kaplan-Yorke dimension has its maximum value of 2.23542.

Now, consider the fractional-order diffusionless Lorenz system given by

$$\begin{cases} \dfrac{d^\alpha x}{dt} = -y - x \\ \dfrac{d^\beta y}{dt} = -xz \\ \dfrac{d^\gamma z}{dt} = xy + R \end{cases}, \qquad (4)$$

where $\alpha, \beta, \gamma$ determine the fractional order, $0 < \alpha, \beta, \gamma \leq 1$. Figure 1(b) shows the chaotic attractor for the fractional-order diffusionless Lorenz system with $\alpha = \beta = \gamma = 0.95$ and $R = 3.4693$.

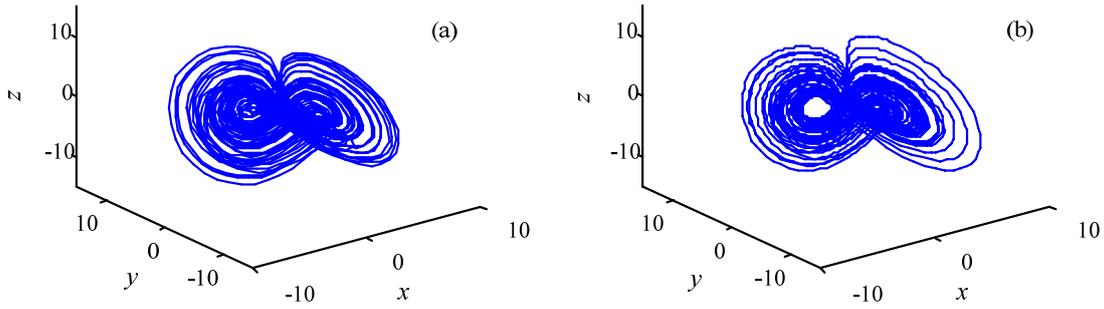

Fig. 1 Chaotic attractors for the diffusionless Lorenz system in Eq. (4) with $R=3.4693$.

(a) $\alpha=\beta=\gamma=1$     (b) $\alpha=\beta=\gamma=0.95$

There are two ways to study fractional-order systems. One is through linear approximations. By using frequency-domain techniques based on Bode diagrams, one can obtain a linear approximation for the fractional-order integrator, the order of which depends on the desired bandwidth and the discrepancy between the actual and approximate Bode diagrams. The other is the Adams-Bashforth-Moulton predictor-corrector scheme [27-29], which is a time-domain approach and thus is more effective. Here, we derive a generalization of the Adams-Bashforth-Moulton scheme appropriate for Eq. (4).

The following differential equation

$$\begin{cases} \dfrac{d^\alpha x}{dt^\alpha} = f(t,x), & 0 \leq t \leq T \\ x^k(0) = x_0^{(k)} & k = 0,1,2,\cdots,\lceil \alpha \rceil - 1 \end{cases}, \qquad (5)$$

is equivalent to the Volterra integral equation [27]

$$x(t)=\sum_{k=0}^{n-1}x_0^{(k)}\frac{t^k}{k!}+\frac{1}{\Gamma(\alpha)}\int_0^t(t-\tau)^{(\alpha-1)}f(\tau,x(t))d\tau. \tag{6}$$

Obviously, the sum on the right-hand side is completely determined by the initial values, and hence is known. In a typical situation, one has $0<\alpha<1$, and hence the Volterra equation (5) is weakly singular. In [27-29], the predictor-corrector scheme for equation (5) is derived, and this approach can be considered to be an analogue of the classical one-step Adams-Moulton algorithm.

Set $h=T/N$ and $t_j=jh(j=0,1,2,\cdots,N)$ with $T$ being the upper bound of the interval on which we are looking for the solution. Then the corrector formula for Eq. (6) is given by

$$x_h(t_{n+1})=\sum_{k=0}^{\lceil\alpha\rceil-1}x_0^{(k)}\frac{t_{n+1}^k}{k!}+\frac{h^\alpha}{\Gamma(\alpha+2)}f(t_{n+1},x_h^p(t_{n+1}))+\frac{h^\alpha}{\Gamma(\alpha+2)}\sum_{j=0}^n a_{j,n+1}f(t_j,x_h(t_j)), \tag{7}$$

where

$$a_{j,n+1}=\begin{cases}n^{\alpha+1}-(n-\alpha)(n+1)^\alpha, & j=0\\(n-j+2)^{\alpha+1}+(n-j)^{\alpha+1}-2(n-j+1)^{\alpha+1} & 1\le j\le n\end{cases}. \tag{8}$$

By using a one-step Adams-Bashforth rule instead of a one-step Adams-Moulton rule, the predictor $x_h^p(t_{n+1})$ is given by

$$x_h^p(t_{n+1})=\sum_{k=0}^{n-1}x_0^k\frac{t_{n+1}^k}{k!}+\frac{1}{\Gamma(\alpha)}\sum_{j=0}^n b_{j,n+1}f(t_j,x_h(t_j)), \tag{9}$$

where

$$b_{j,n+1}=\frac{h^\alpha}{\alpha}((n-j+1)^\alpha-(n-j)^\alpha),\ 0\le j\le n. \tag{10}$$

Now, the basic algorithm for the fractional Adams-Bashforth-Moulton method is completely described by Eqs. (7) and (9) with the weights $a_{j,n+1}$ and $b_{j,n+1}$ being defined according to (8) and (10), respectively.

The error estimate of this method is

$$e=\max_{j=0,1,\cdots,N}|x(t_j)-x_h(t_j)|=O(h^p) \tag{11}$$

where $p=\min(2,1+\alpha)$. Using this method, the fractional-order diffusionless Lorenz equations (4) can be written as

$$\begin{cases}x_{n+1}=x_0+\dfrac{h^\alpha}{\Gamma(\alpha+2)}\{[-y_{n+1}^p-x_{n+1}^p]+\sum_{j=0}^n a_{1,j,n+1}(-y_j-x_j)\}\\[6pt] y_{n+1}=y_0+\dfrac{h^\beta}{\Gamma(\beta+2)}[(-x_{n+1}^p z_{n+1}^p)+\sum_{j=0}^n a_{2,j,n+1}(-x_j z_j)]\\[6pt] z_{n+1}=z_0+\dfrac{h^\gamma}{\Gamma(\gamma+2)}\{[x_{n+1}^p y_{n+1}^p+R]+\sum_{j=0}^n a_{3,j,n+1}(x_j y_j+R)\}\end{cases}, \tag{12}$$

where

$$\begin{cases} x_{n+1}^p = x_0 + \dfrac{1}{\Gamma(\alpha)}\sum_{j=0}^{n} b_{1,j,n+1}(-y_j - x_j) \\ y_{n+1}^p = y_0 + \dfrac{1}{\Gamma(\beta)}\sum_{j=0}^{n} b_{2,j,n+1}(-x_j z_j) \\ z_{n+1}^p = z_0 + \dfrac{1}{\Gamma(\gamma)}\sum_{j=0}^{n} b_{3,j,n+1}(x_j y_j + R) \end{cases},$$

$$\begin{cases} b_{1,j,n+1} = \dfrac{h^{\alpha}}{\alpha}((n-j+1)^{\alpha} - (n-j)^{\alpha}), & 0 \le j \le n \\ b_{2,j,n+1} = \dfrac{h^{\beta}}{\beta}((n-j+1)^{\beta} - (n-j)^{\beta}), & 0 \le j \le n, \\ b_{3,j,n+1} = \dfrac{h^{\gamma}}{\gamma}((n-j+1)^{\gamma} - (n-j)^{\gamma}), & 0 \le j \le n \end{cases}$$

$$\begin{cases} a_{1,j,n+1} = \begin{cases} n^{\alpha} - (n-\alpha)(n+1)^{\alpha} & j = 0 \\ (n-j+2)^{\alpha+1} + (n-j)^{\alpha+1} - 2(n-j+1)^{\alpha+1}, & 1 \le j \le n \end{cases} \\ a_{2,j,n+1} = \begin{cases} n^{\beta} - (n-\beta)(n+1)^{\beta} & j = 0 \\ (n-j+2)^{\beta+1} + (n-j)^{\beta+1} - 2(n-j+1)^{\beta+1}, & 1 \le j \le n \end{cases} \\ a_{3,j,n+1} = \begin{cases} n^{\gamma} - (n-\gamma)(n+1)^{\gamma} & j = 0 \\ (n-j+2)^{\gamma+1} + (n-j)^{\gamma+1} - 2(n-j+1)^{\gamma+1}, & 1 \le j \le n \end{cases} \end{cases}.$$

## 3. Bifurcations of the fractional-order diffusionless Lorenz system

### 3.1. Fractional-order diffusionless Lorenz chaotic system

In our simulations, we have visually inspected the bifurcation diagrams to identify chaos. We also have confirmed these by calculating the largest Lyapunov exponent in some cases using the Wolf algorithm [30]. Here, we let $\alpha=\beta=\gamma=q$, and the effective chaotic range of the fractional-order diffusionless system with different control parameter is found as shown in Fig. 2. Clearly there exist three different states corresponding to limit cycles, chaos, and convergence. If the integer-order system is chaotic, the effective chaotic range of the fractional-order system decreases slightly as control parameter increases. If the integer order system is a limit cycle, the fractional-order system can produce chaos, but the effective chaotic range of the fractional-order system shrinks significantly as the control parameter increases. The largest Lyapunov exponent of the fractional-order system with $R=8$ is shown in Fig. 3. It is consistent with the chaotic range at this value.

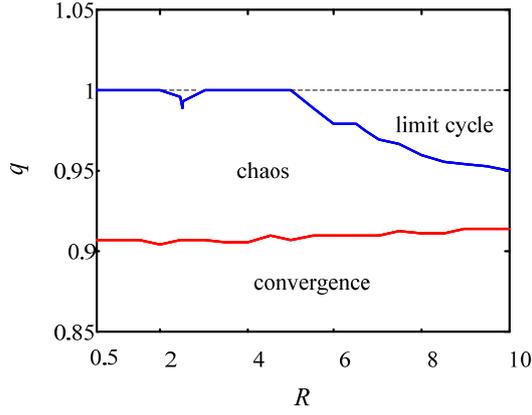 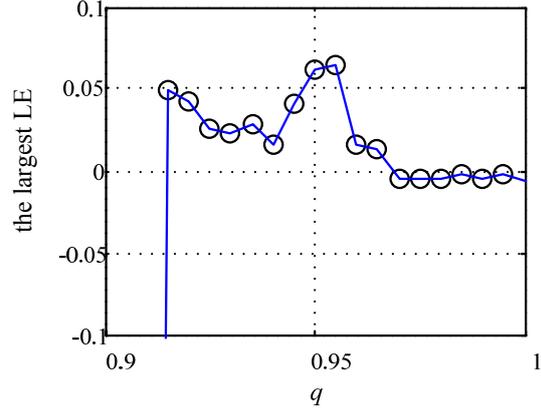

Fig. 2 Chaotic range of the fractional order DLE.  Fig. 3 Largest LE of the fractional order DLE with $R$=8.

### 3.2. Bifurcations with different control parameter $R$

Here, the fractional orders $\alpha$, $\beta$, $\gamma$ are equal and fixed at 0.95 while the control parameter $R$ is varied from 0.5 to 11. The initial states of the fractional-order diffusionless Lorenz system are $x(0) = -0.0249$, $y(0) = -0.1563$, and $z(0) = 0.9441$. For a step size in $R$ is 0.01 and the running time is 140s, the bifurcation diagram in Fig. 4 was obtained. It shows that the fractional-order DLE is chaotic with one periodic widow when the total order is $\alpha = \beta = \gamma = 2.85$. When the control parameter $R$ is decreased from 11, the fractional-order system enters into chaos by a period-doubling bifurcation as shown in Fig. 5(a) (with steps of 0.005). To observe the dynamic behavior, the periodic window is expanded with steps of 0.005, as shown in Fig. 5(b). There exists an interior crisis when $R \cong 5.57$, a flip bifurcation when $R \cong 6.02$, and a tangent bifurcation when $R \cong 6.55$. In the interior crisis, the chaotic attractor collides with an unstable periodic orbit or limit cycle within its basin of attraction. When the collision occurs, the attractor suddenly expands in size but remains bounded. For the tangent bifurcation, a saddle point and a stable node coalesce and annihilate one another, producing an orbit that has periods of chaos interspersed with periods of regular oscillation [31]. This same behavior occurs in the periodic windows of the logistic map, including the miniature windows within the larger windows [32]. In this periodic window, we also observe the route to chaos by a period-doubling bifurcation.

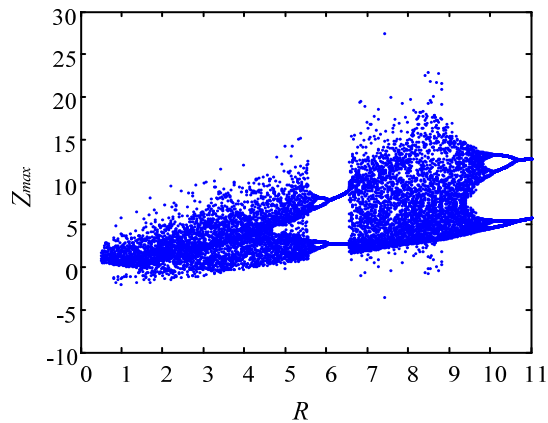

Fig. 4 Bifurcation diagram of the fractional-order diffusionless Lorenz system with $R$ for $q$=0.95.

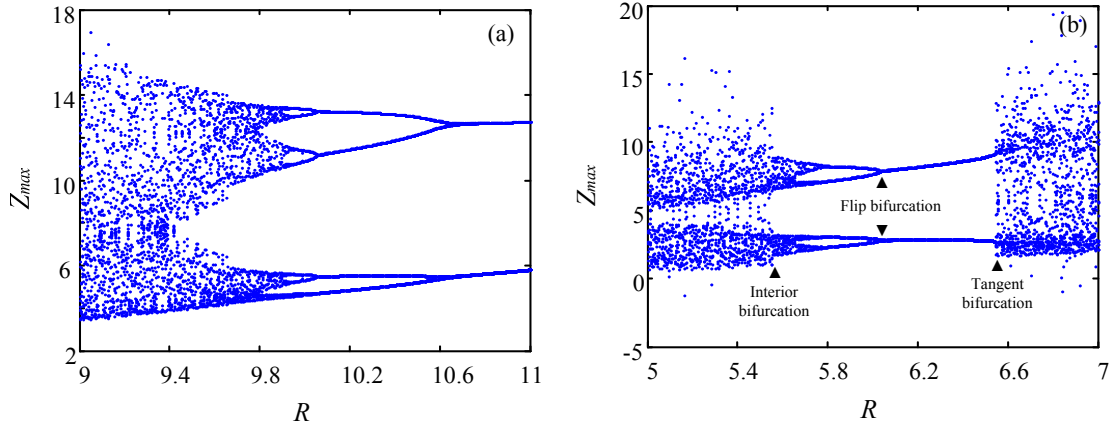

Fig. 5 Bifurcation diagram of the fractional-order diffusionless Lorenz system with $R$ for $q$=0.95.
(a) $R \in [9,11]$  (b) $R \in [5,7]$

### 3.3. Bifurcations with different fractional orders

Now let $\alpha=\beta=\gamma=q$, and change the fractional order $q$ from 0.9 to 1, but fix the control parameter at $R$=8. The initial states of the fractional-order diffusionless Lorenz system and running time kept the same as above, but with a variational step of $R$ set to 0.0005 gives the bifurcation diagram shown in Fig. 6. This case shows the route to chaos for the fractional-order DLE as the fractional order decreases. It is interesting to note that a chaotic transient is observed when $q$ is less than 0.912. The state space trajectory is shown in Fig. 7(a) for $q$=0.911, which suddenly switches to a pattern of oscillation that decays to the right equilibrium point. The time history of $z(t)$ in Fig. 7(b) also shows eventual convergence to the fixed point. On the average, chaotic behavior switches to damped behavior after about 70 oscillations. For larger $q < q_0 \approx 0.912$, chaotic behavior persists longer. Similar behavior has been reported for the standard Lorenz system [33]. The fractional-order system gives way to chaos by a period-doubling bifurcation as shown in Fig. 8(a) with steps in $R$ of 0.0002. It is chaotic in the range of 0.92 to 0.962. When the fractional order is less than 0.92, the fractional-order system converges to a fixed point. Expanding the periodic window shows the dynamic behavior better as in Fig. 8(b) with steps of 0.0001. There exist three kinds of bifurcation, i.c., a tangent bifurcation, a flip bifurcation, and an interior crisis. The fractional order parameter can be taken as a bifurcation parameter, just like the control parameter.

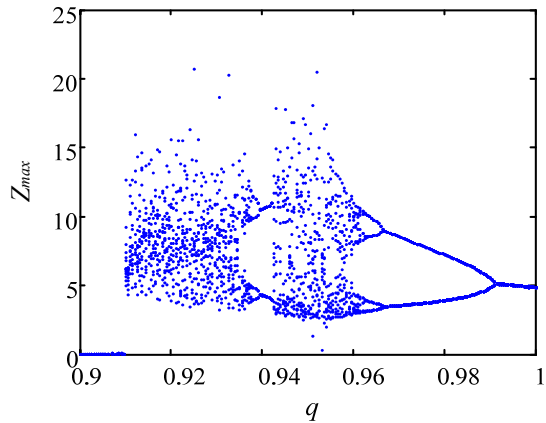

Fig. 6 Bifurcation diagram of the fractional-order diffusionless Lorenz system with $q$ for $R$=8.

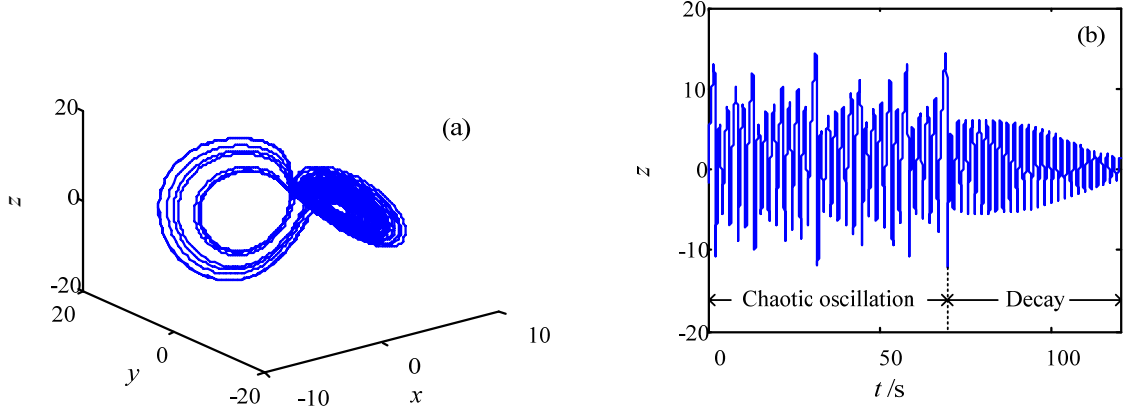

Fig. 7 Transient chaotic behavior in system (4) with $R=8$ and $q=0.911$.
(a) 3D view on the $x$-$y$-$z$ space    (b) The time histories of variable $z$

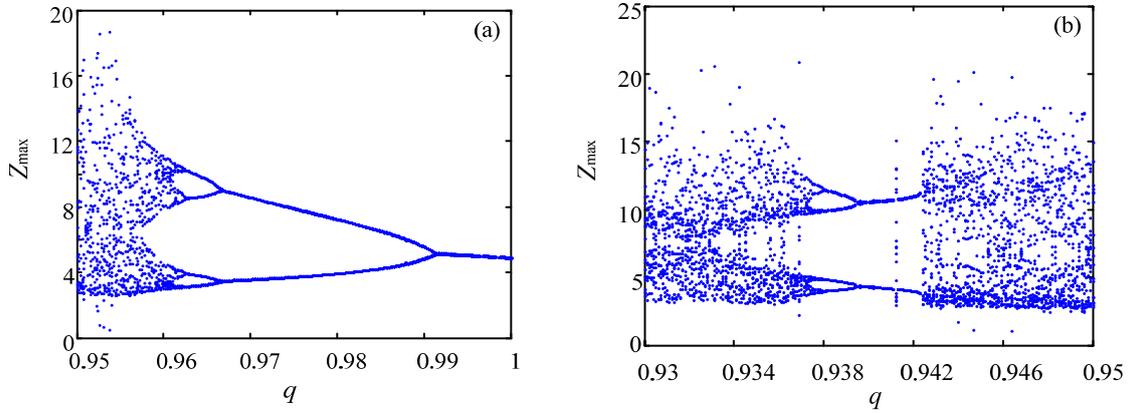

Fig. 8 Bifurcation diagram of the fractional-order diffusionless Lorenz system with $q$ for $R=8$.
(a) $q \in [0.95,1]$  (b) $q \in [0.93,0.95]$

### 3.4. Bifurcations with different fractional order for the three equations

(1) Fix $\beta=\gamma=1$, $R=8$, and let $\alpha$ vary. The system is calculated numerically for $\alpha \in [0.4,1]$ with an increment of $\alpha$ equal to 0.002. The bifurcation diagram is shown in Fig. 9(a). It is found that when $0.43 \leq \alpha \leq 0.94$, the fractional-order system is chaotic with one periodic window at $c \in (0.484, 0.503)$ as shown in Fig. 9(b) with steps of 0.0002. When $\alpha$ increases from 0.4, or decreases from 1, a period-doubling route to chaos is observed.

(2) Fix $\alpha=\gamma=1$, $R=8$, and let $\beta$ vary. The system is calculated numerically for $\beta \in [0.75,1]$ with an increment of $\beta$ equal to 0.001. The bifurcation diagram is shown in Fig. 10. When $\beta$ decreases from 1, a period-doubling route to chaos is observed, and it converges to a fixed point when $\beta$ is less than 0.8. Thus the total smallest order of the fractional-order diffusionless Lorenz chaotic system is 2.8 in this case.

(3) Fix $\alpha=\beta=1$, $R=8$, and let $\gamma$ vary. The system is calculated numerically for $\gamma \in [0.8,1]$ with an increment of $\gamma$ equal to 0.001. The bifurcation diagram is shown in Fig. 11(a). Similar phenomena are found, but the chaotic range of the fractional-order system is much smaller than that of the previous two cases, and it enters into chaos by a period-doubling route, and then it converges to a fixed point when $\gamma=0.816$. The dynamic behaviors in the periodic window are similar with the case of (2) as shown in Fig. 11(b) with steps of 0.0002.

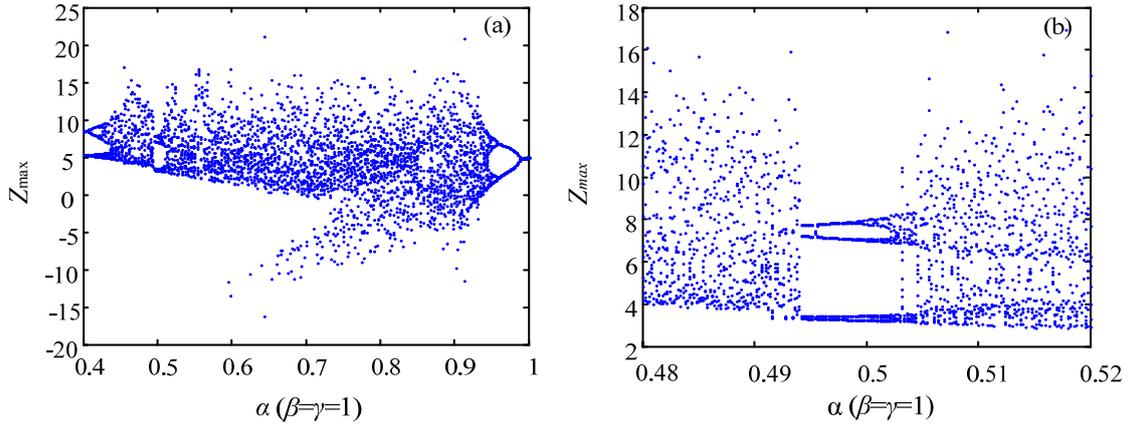

Fig. 9 Bifurcation diagram of the fractional-order diffusionless Lorenz system with α for β=γ=1.
(a) $\alpha \in [0.4,1]$  (b) $\alpha \in [0.48,0.52]$

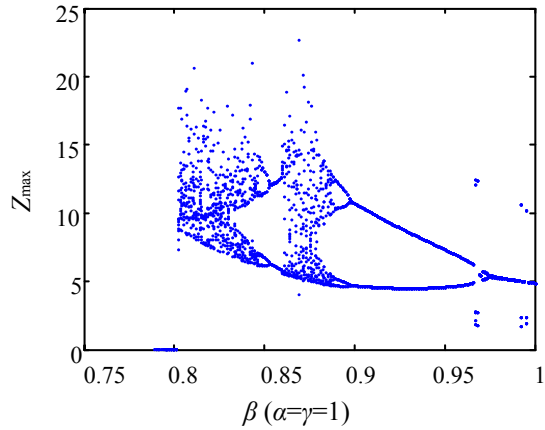

Fig. 10 Bifurcation diagram of the fractional-order diffusionless Lorenz system with β for α=γ=1.

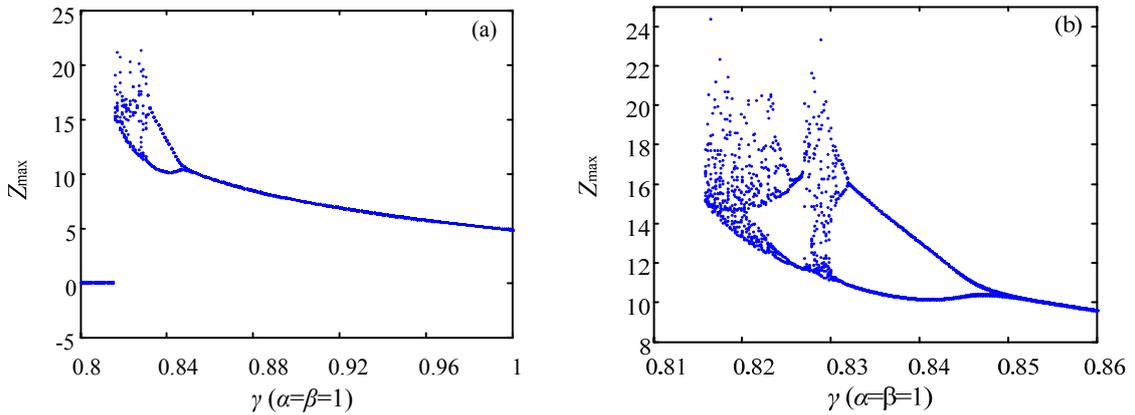

Fig. 11 Bifurcation diagram of the fractional-order diffusionless Lorenz system with γ for α=β=1.
(a) $\gamma \in [0.8,1]$  (b) $\gamma \in [0.81,86]$

## 4. Conclusions

In this paper, we have numerically studied the bifurcations and dynamics of the fractional-order diffusionless Lorenz system by varying the system parameter and the system order. Typical bifurcations such as period-doubling bifurcations, flip bifurcations, tangent bifurcations, and interior crisis bifurcations were observed when both the control parameter and the fractional-order are changed. Complex dynamic behaviors such as fixed point, periodic motion, transient chaos, and chaos, occur in this fractional-order system. Future work on the topic should include a theoretical analysis of the dynamics of the fractional-order system, as well as in-depth studies of chaos control and synchronization for the system.

## Acknowledgments

This work was supported by the China Scholarship Council (No.2006A39010), the National Nature Science Foundation of People's Republic of China (Grant No.60672041), and the National Science Foundation for Post-doctoral Scientists of People's Republic of China (Grant No. 20070420774). We are grateful for discussions with Prof. George Rowlands.